# *Properties of amorphous CoZr(RE) (RE=Gd, Sm, Dy) films with various uniaxial anisotropies, prepared by a new process*

D.H. Shin [a,*], H. Niedoba[b], Y. Henry[c], F. Machizaud[d], V. Brien[d], D. Chumakov[e], R. Schafer[e], G. Suran[a]

a Laboratoire Louis Néel, CNRS, B.P.166 Grenoble 38042, Cedex, France
b Laboratoire de Magnétisme et d'Optique, CNRS, Université de Versailles, Versailles 78035 Cedex, France
c Institut de Physique et Chimie, IPCMS, Strasbourg 67037, France
d Laboratoire de Science et Genie des Matériaux, EMN, Nancy 54042 France
e Institut für Festkorper und Werkstofforschung P.O. Box 270016, Dresden 01069 Germany
*Corresponding author.

**Abstract**
***If a DC magnetic field is applied parallel to the plane of amorphous CoZr(RE) thin films during sputter depositing, uniaxial anisotropy is formed the direction of which depends upon the choice of RE substituted and its concentration. When RE = Gd a perpendicular anisotropy Kp forms over a large concentration range, a spin reorientation process being at the origin of the process. A well-defined Kp is developed also in CoFeZrGd and CoZrGdSm films. CoZrGdDy films exhibit simultaneously a perpendicular and an in-plane uniaxial anisotropy. The related magnetization process and domain structures are quite peculiar***.

## 1. *Introduction*

Amorphous transition metal (TM)–rare earth (RE) (a-(TM)–(RE)) thin films were studied very intensively [1] since the discovery in 1973 that they can exhibit a perpendicular uniaxial anisotropy ($K_p$), and are promising materials for magneto-optic recording devices. Frequently the technical problem has been to achieve a-RE–TM films which exhibit a high perpendicular anisotropy field $H_p = 2K_p/M_s$. The preparation was performed by sputtering or evaporation, but it was necessary to use specific deposition conditions. If one wanted to produce $H_p$ in sputter deposited films, one had to apply a negative DC bias voltage ($-V_b$). The $H_p$ increased for higher $-V_b$ meanwhile the magnetic homogeneity degraded. Only a limited range of compounds with large enough $H_p$ could be obtained by evaporation. Clearly, any new deposition process is welcome if the films have interesting properties. In the present work, we report the principle and the various potentialities of a new preparation technique. It is based upon an investigation, that if a DC magnetic field $H_{DC}$ is applied in the plane of an a-$Co_{95-x}Zr_5(RE)_x$ film during sputter deposition, a uniaxial anisotropy is formed the direction of which depend upon the RE substituted, a study made for a low RE content ($0 < x \leq 4$) [2]. When the RE was Pr, Nd, Dy or Tb, an in-plane uniaxial anisotropy $K_u$ developed with its easy axis parallel to $H_{DC}$. However, when the RE was Gd or Sm a highly unusual effect occurred. In this case the formation of a coherent anisotropy perpendicular to the direction of the $H_{DC}$ field, so a uniaxial anisotropy perpendicular to the film plane $K_p$ was observed.

The present investigation is based upon the specific behavior of a-TM–RE thin films containing Gd. We report the properties of a-GdCoZr films for a large concentration range. The effects of substituted Fe or Sm upon the properties of a-GdCoZr are summarized. When a RE was substituted which favors the formation of $K_u$; films, which exhibit both $K_p$ and $K_u$ were obtained, as shown by the study of CoZrGdDy films. An original behavior of the magnetization process and a peculiar domain structure were observed for a particular content of the two RE.

### *2. Experimental procedure*

The a-TM–RE thin films were prepared by RF diode sputtering with Ar as sputter gas. The $Co_{95}Zr_5$ target was stuck to the cathode, on which triangular shaped RE(Gd, Sm, Dy) and Zr platelets were positioned. The thickness t of the as studied films was typically t = 250 ± 50 nm. Our deposition conditions differ fundamentally from those used classically. Generally during deposition, a direct contact exists between the film and the plasma. Consequently, it is difficult to control the real deposition temperature, and the formation of the deposit is strongly perturbed by the Argon ions which hit the surface of the film. We used a substrate holder having a specific configuration [3]. It is composed of a couple of permanent magnets which generate $H_{DC}$ ~ 1000 Oe, recovered by a Fe plate, the deposition occurs through a slit in its centre. Consequently, the magnetic field is strictly parallel to the substrate. The plasma is established between the target and the Fe plate, there is no direct contact between the plasma and the substrate, and the effect of Ar ion bombardment is largely reduced. We verified in experiments that the use of this substrate holder, where the substrate is kept outside the plasma, is necessary in order to obtain the present results. Experiments, displayed mainly on semiconductor films, showed that the properties of a film deposited outside the plasma differs strongly from samples prepared using the classical configuration.

### *3. Experimental results*
### *3.1. CoZrGd, CoZrGdFe and CoZrGdSm films*

The magnetic and structural properties of $Co_{95_x}Zr_5Gd_x$ films for 7<x<30 was studied in detail. We deduced both $K_p$ and saturation magnetization $M_s$ from torque measurements [4]. The compositional variation of $M_s$; measured at room temperature, was similar to that obtained previously [5]. We determined the variation of $K_p$ as a function of the Gd content, and the deposition parameters. $K_p$ varies essentially as a function of the pressure of the sputter gas $P_{Ar}$. The variation of $K_p$ is fairly spectacular for the concentration range where compensation temperature ($T_{comp.}$) is close to room temperature. Study of films with x = 22.7 at% showed that $K_p$ exhibits a well-defined maximum for $P_{Ar}$ = 5 mTorr. When $P_{Ar}$ was too high, e.g. 15mTorr, an in-plane anisotropy formed essentially (Fig. 1). The compositional variation of $K_p$; determined for $P_{Ar}$ = 5 mTorr, is reported in Fig. 2. $K_p$ is maximum for x close to x ~ 10; and then decreases slowly but continuously up to x ~ 24: Films rich in Gd (x > 28 at%) exhibited an in-plane anisotropy. The quality factor $Q = K_p/ 2\Pi M_s^2$ is higher than 1 for a quite large concentration range (Fig. 2). This compositional dependence of $K_p$ is rather unusual, in various binary a-TM–REsystems the maximal of $K_p$ was found for 20–40 at% RE [1]. The present result is probably related to the particular mechanism of formation of $K_p$ which is a specific one. A study of the hysteresis loops measured by polar Kerr magnetometry revealed a spin reorientation process (Fig. 3). On films with x = 19-25% a rectangular loop, typical of $K_p$; was recorded at room temperature which transformed progressively to a loop with in-plane anisotropy as the measuring temperature was raised, while for x > 28 at% the opposite process occurred. Such a change of the magnetization direction with temperature is observed for the first time in a single phase amorphous film [6]. The dependence upon composition of the reorientation process explains the variations of $K_u$: $K_u$ is fairly small with $K_u$5$K_p$ for xo22: When 23oxo25 $K_p$ still dominates the magnetization process but the in-plane component of the magnetization is no more negligible. The spin reorientation is the dominant process, because the deposition temperature is controlled carefully. Consequently, the real temperature at which the film forms is determined by the energy of the sputtered particles, thus by $P_{Ar}$.

The structures of the films were characterized using grazing incidence X-ray diffraction. The broad and modulated maximum confirmed the amorphous state. Its deconvolution into five maxima allows one to correlate them to atomic pairs Gd–Gd (one type), Gd–Co (two types), Co–Co (two types). These interatomic distances agree with atomic radii of atomic species and/or atomic pairs of the cluster model [3,7] where the local environment of the RE is the same in the amorphous and the crystalline state as in hexagonal $Co_5RE$, hexagonal $Co_{17}RE_2$, and rhombohedral $Co_{17}RE_2$ structures.

One of the main interests of this deposition process is that films with fairly uniform properties could be obtained. On films with Gd content x ~ 22; a well-defined $K_p$ was obtained for t down to ~ 30 nm, allowing a study of the variation of the magnetic properties as a function of films'thickness.

Substitution of Fe had interesting effects. The local magnetic anisotropy $K_l$ of the Fe is smaller than that of Co, which can explain why $H_c$ decreased as compared to films which contained only Co as TM. The temperature reorientation process from perpendicular to in-plane is also significantly reduced. However, $H_p > 4\pi M_s$ is conserved only for a limited range of Fe up to about 10–15 at%.

Sm possess a large single ion anisotropy so its' substitution should result in an increase of $K_p$ and $H_c$: However, the respective concentration of Sm and Gd in $Co_{95_x}Zr_5Gd_xSm_y$ films have to be carefully adjusted. One must take into account the fact that Sm couples ferromagnetically with Co and one must also limit the amount of the overall RE content to avoid Sm and d as nearest neighbors. For example, large $K_p$ and $H_c$ were observed on samples with Gd around x ~ 19 and Sm y ~ 3-5 content.

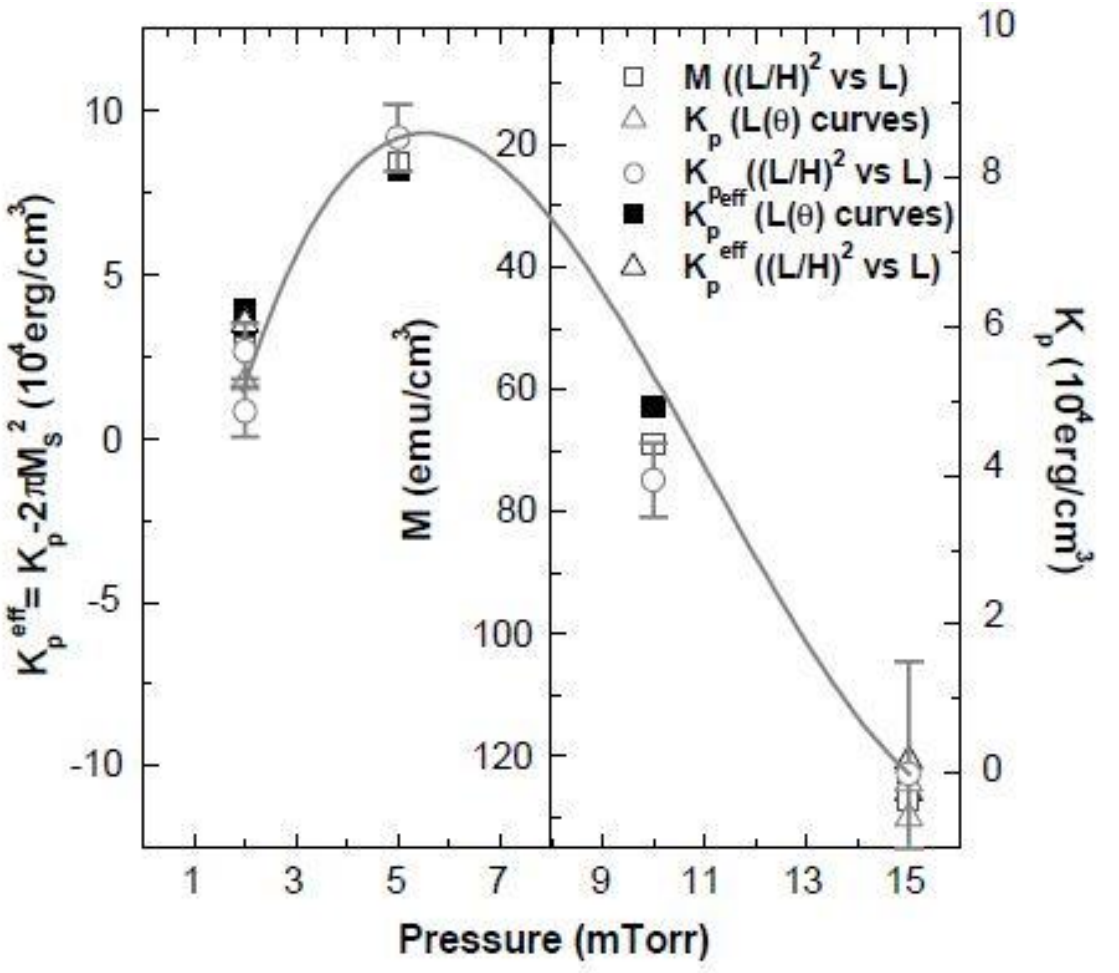

Fig. 1. Variation of the effective perpendicular anisotropy $K_p$ as a function of the sputtering pressure $P_{Ar}$ for $Co_{72.3}Zr_5Gd_{22.7}$ film.

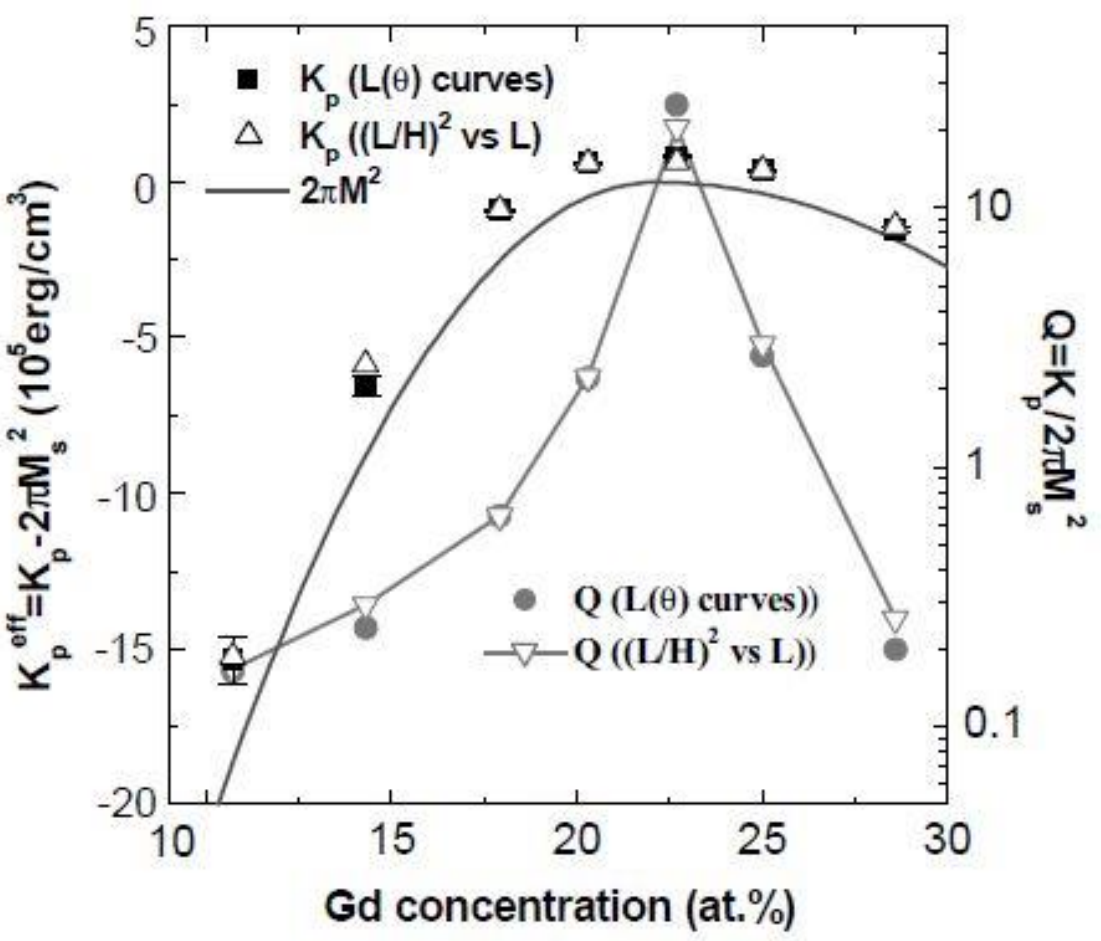

Fig. 2. Effective perpendicular magnetic anisotropy, from torque measurements, and quality factor $Q = K_p/2\pi M_s^2$ vs. composition for CoZrGd films.

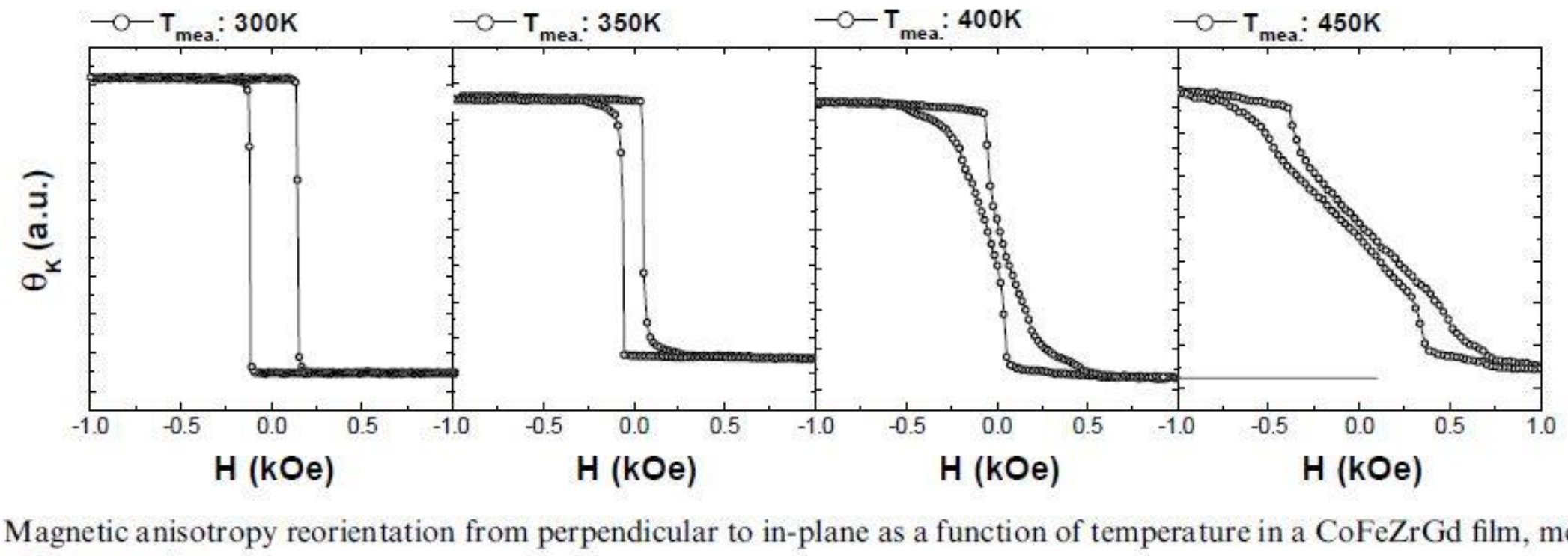

Fig. 3. Magnetic anisotropy reorientation from perpendicular to in-plane as a function of temperature in a CoFeZrGd film, measured by polar Kerr rotation.

### *3.2. CoZrGdDy films*

Films with an original magnetic structure were obtained when two RE, one which favors $K_p$ and the other $K_u$; were substituted simultaneously in a- CoZr. These thin films exhibit two easy axes, one along the film normal and the other along the film plane parallel to $H_{DC}$ as revealed a study performed on CoZrGdDy films. The respective magnitude of $K_p$ and $K_u$—which are related to clusters formed around the Gd and the Dy—could be adjusted by changing respective concentrations of the two RE.

We studied the properties by torque and hysteresis loop measurements. Torque measurements confirmed clearly the existence of $K_p$ and $K_u$ but the data could not be explained by an anisotropy energy of the form $E_p + E_u$: Certainly one has to take into account the fact that the clusters which exhibit $K_p$ and $K_u$ are exchange coupled, but actually a theoretical model does not exist. So the magnitude of $H_p$ and $H_u$ were only estimated.

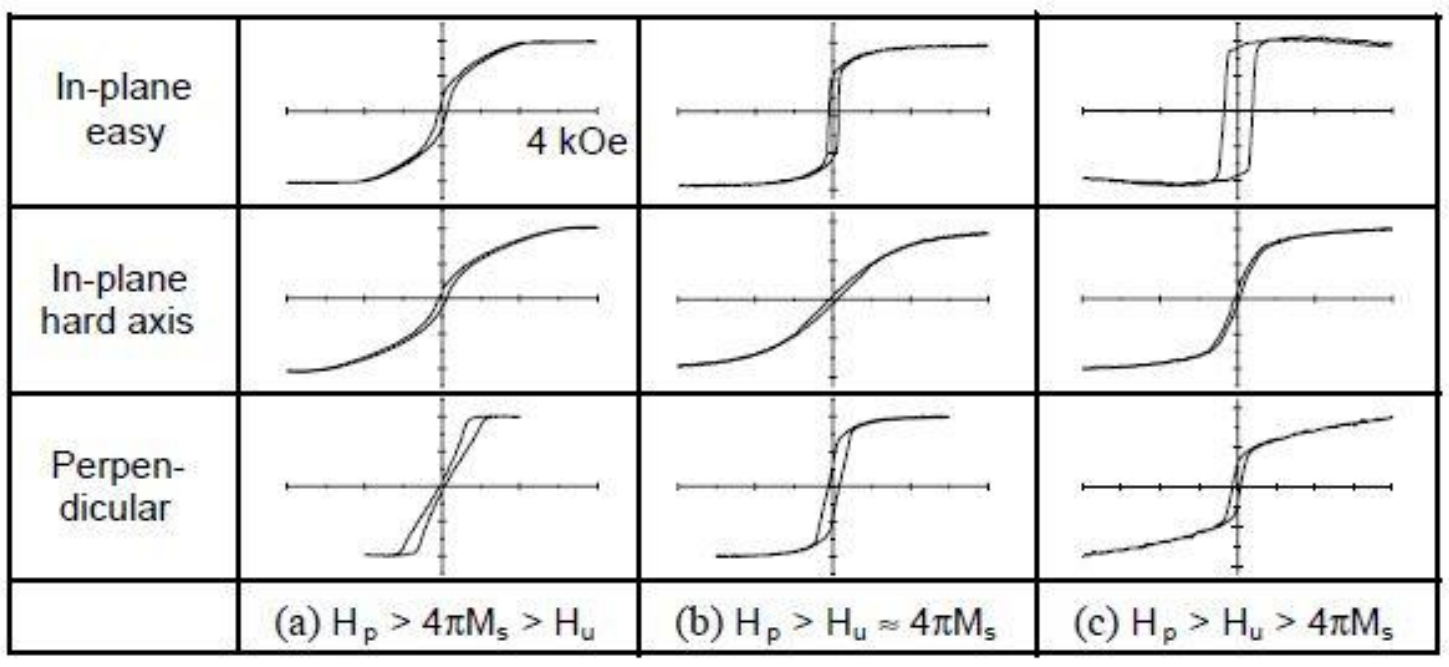

Fig. 4. Magnetization process, measured by VSM at 300 K, of a set of CoZrGdDy films which exhibit both a perpendicular and in-plane uniaxial anisotropy. *M–H* loops for (a) $Co_{73.05}Zr_{4.83}Gd_{13.11}Dy_{9.01}$, (b) $Co_{71.63}Zr_{4.00}Gd_{15.31}Dy_{9.04}$, and (c) $Co_{69.55}Zr_{4.94}Gd_{16.49}$-$Dy_{9.02}$.

The study of the magnetization process along the in-plane easy and hard axis and along the film normal showed that the shape of the loops and $T_{comp.}$ changed rapidly even for relatively small changes of the RE concentration. An original magnetization process was observed on films with $T_{comp.}$ ~ 300K and $H_p > H_u > 4\pi M_s$ : a loop close to a rectangular one was detected along the in-plane easy axis and the film normal, while the shape of the loop along the in-plane hard axis was a classical one, exhibiting essentially $K_u$ (Fig. 4).

The structure and the behavior of the domains were studied on a set of films having the composition around $Co_{73\pm x\pm y}Zr_5Gd_{14\pm x}Dy_{8\pm y}$ with x ≈ 2 and y ≈ 1. The change of the RE concentration by this small amount weakly affected the respective magnitude of $H_p$ and $H_u$—we always had $H_p > H_u$—while $4\pi M_s$ varied strongly as shown also by the variation of Tcomp. The overall behavior of the domains are determined by the respective magnitudes of $H_p$, $H_u$ and $4\pi M_s$. On a film with quite low $T_{comp.}$—where one has $H_p > 4\pi M_s > H_u$—a regular stripe domain structure was found exhibiting a classical behavior. After demagnetization a mixed pattern consisting of stripes and bubbles are observed. In a field perpendicular to the film surface a stripe nucleation and growth occurs similar to a classical one, results which show that the effect of $H_u$ upon the process is small. On a film with $T_{comp.}$ ~ 240 K—so $H_p > 4\pi M_s \sim H_u$—stripe type domains are observed but with a behavior strongly influenced by $H_u$ (Fig. 5). When the field is applied along the in-plane easy axis a complete reversal process is observed, the nucleation and growing into irregular dot. For field applied along the in-plane hard axis the stripe domain structure still appears but it does not follow the direction of the applied field and seems to develop by some rotational process. The field required for saturation is an order of magnitude higher than for the case corresponding to the easy axis.

Samples which exhibit a $T_{comp.}$ ~ 300K—so when $H_p > H_u > 4\pi M_s$—exhibited a fairly unusual behavior. In this case two independent kinds of domains and magnetization processes can be observed. Wide and irregular out-of -plane domains, are observed which can be modified only by magnetic fields perpendicular to the surface. Simultaneously, there is an in-plane structure that can be switched by relatively low in-plane fields. The full results of this investigation will be reported elsewhere.

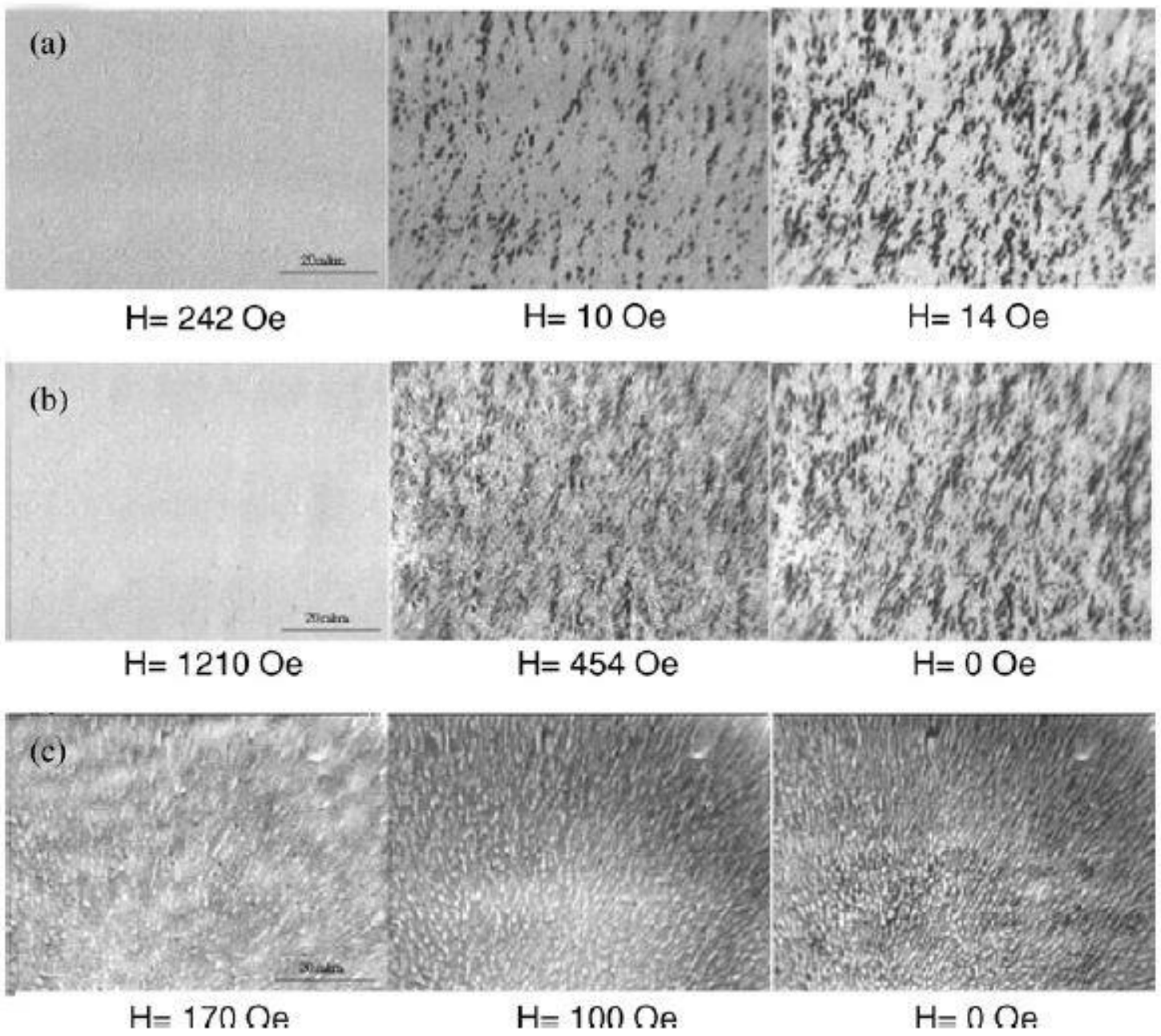

Fig. 5. Change of the domain structure as a function of the magnitude and the orientation of the applied field $H$ in a CoZrGdDy film where $H_p > H_u \sim 4\pi M_s$. $H$ applied along (a) the in-plane easy axis (b) the in-plane hard axis (c) the film normal.

## *4. Conclusions*

It was shown that in CoGd, CoFeGd or CoSmGd films for a range of composition a perpendicular anisotropy field forms, if a DC magnetic field is applied parallel to the film plane during the preparation by sputtering. A spin reorientation mechanism is the origin of this process. a-CoGd(RE) films exhibit simultaneously $K_p$ and $K_u$—if an RE such as Dy which favors the formation of an in-plane uniaxial anisotropy is substituted. The respective values of $K_p$ and $K_u$ can be adjusted by changing the composition. These films exhibit an original domain structure the investigation of which is in progress.

## *References*

[1] P. Hansen, C. Clausen, G. Much, M. Rosenkratz, K. Witter, J. Appl. Phys. 66 (1989) 756.
[2] G. Suran, H. Ouahmane, M. Rivoire, J. Sztern, J. Appl.Phys. 73 (1993) 5721.
[3] D.H. Shin, H. Niedoba, G. Suran, IEEE Trans. Magn, to appear.
[4] H. Miyajima, K. Sato, T. Mizoguchi, J. Appl. Phys. 47 (1976) 4669.
[5] R.C. Taylor, A. Gangulee, J. Appl. Phys. 47 (1976) 4666.
[6] E. Stravrou, K. R. oll, J. Appl. Phys. 85 (1999) 5971.
[7] G. Suran, F. Machizaud, IEEE Trans. Magn. 36 (2000) 2924.